\documentclass[pra,twocolumn]{revtex4}
\usepackage{amsmath}
\usepackage{amsfonts}
\usepackage{amssymb}
\usepackage{graphicx}
\usepackage{sidecap}
\usepackage{caption,subcaption}
\usepackage{bm}

\usepackage{graphicx}
\newcommand{\graphr}{
\setlength{\unitlength}{1.8ex}
\begin{picture}(1.05,1)
\linethickness{1pt}
\put(.5,0){\circle*{.2}}
\put(0.5,.5){\circle{1}}
\end{picture}
}

\newcommand{\graphgamma}{
\setlength{\unitlength}{1.8ex}
\begin{picture}(1,1)(0,-0.25)
\linethickness{.5pt}
\put(0,0){\line(1,0){1}}
\put(0,0){\circle*{.2}}
\put(1,0){\circle*{.2}}
\end{picture}
}

\newcommand{\graphrra}{
\setlength{\unitlength}{1.8ex}
\begin{picture}(2,.5)(-.5,-0.25)
\put(0.5,0){\circle*{.2}}
\linethickness{.5pt}
\put(0,0){\circle{1}}
\put(1,0){\circle{1}}
\end{picture}
}

\newcommand{\graphrrb}{
\setlength{\unitlength}{1.8ex}
\begin{picture}(3,.5)(-0.25,.1)
\linethickness{1pt}
\put(0.50,0.00){\circle*{.2}}
\put(0.50,0.50){\circle{1}}

\put(2,0.00){\circle*{.2}}
\put(2,0.50){\circle{1}}

\end{picture}
}

\newcommand{\graphgra}{
\setlength{\unitlength}{1.8ex}
\begin{picture}(2.5,.5)(-0.25,-0.25)
\put(1,0){\circle*{.2}}
\put(2,0){\circle*{.2}}
\linethickness{1pt}
\put(0.5,0){\circle{1}}
\linethickness{.5pt}
\put(1,0){\line(1,0){1}}
\end{picture}
}

\newcommand{\graphgrb}{
\setlength{\unitlength}{1.8ex}
\begin{picture}(3.5,.5)(-0.25,-0.25)
\linethickness{1pt}
\put(1,0){\circle*{.2}}
\put(2,0){\circle*{.2}}
\put(3,0){\circle*{.2}}
\put(0.5,0){\circle{1}}

\linethickness{.5pt}
\put(2,0){\line(1,0){1}}
\end{picture}
}

\newcommand{\graphgga}{
\setlength{\unitlength}{1.8ex}
\begin{picture}(1.5,.5)(-0.25,-0.25)
\linethickness{.5pt}
\put(0,0){\circle*{.2}}
\put(1,0){\circle*{.2}}
\linethickness{1pt}
\put(0.5,0){\circle{1}}
\end{picture}
}

\newcommand{\graphggb}{
\setlength{\unitlength}{1.8ex}
\begin{picture}(1.5,.5)(-0.25,-0.25)
\linethickness{.5pt}
\put(0,-0.5){\circle*{.2}}
\put(0,-0.5){\line(1,0){1}}
\put(1,-.5){\circle*{.2}}
\put(1,-0.5){\line(0,1){1}}
\put(1,0.5){\circle*{.2}}
\end{picture}
}

\newcommand{\graphggc}{
\setlength{\unitlength}{1.8ex}
\begin{picture}(1.5,.5)(-0.25,-0.25)
\linethickness{.5pt}
\put(0,0.5){\line(1,0){1}}
\put(0,-0.5){\line(1,0){1}}
\linethickness{.5pt}
\put(0,0.5){\circle*{0.2}}
\put(0,-0.5){\circle*{0.2}}
\put(1,0.5){\circle*{0.2}}
\put(1,-0.5){\circle*{0.2}}
\end{picture}
}

\newcommand{\graphrrra}{
\setlength{\unitlength}{1.5ex}
\begin{picture}(3,1.5)(-1.5,-0.5)
\linethickness{0.001em}
\put(0,0){\circle*{0.5}}
\qbezier(0.00,0.00)(0.281,1.875)(1.125,0.750)
\qbezier(1.125,0.750)(1.969,-0.375)(0.00,0.00)
\qbezier(0.00,0.00)(-1.406,-1.125)(0.00,-1.125)
\qbezier(0.00,-1.125)(1.406,-1.125)(0.00,0.00)
\qbezier(0.00,0.00)(-1.969,-0.375)(-1.125,0.750)
\qbezier(-1.125,0.750)(-0.281,1.875)(0.00,0.00)
\end{picture}}

\newcommand{\graphrrrb}{
\setlength{\unitlength}{1.8ex}
\begin{picture}(3,1)(-0.25,-0.25)
\linethickness{1pt}
\put(0.5,0){\circle*{.2}}
\linethickness{1pt}
\put(2,0){\circle*{.2}}
\put(0.5,-0.5){\circle{1}}
\put(0.5,0.5){\circle{1}}
\put(2,0.5){\circle{1}}
\end{picture}
}

\newcommand{\graphrrrc}{
\setlength{\unitlength}{1.8ex}
\begin{picture}(4.50,1)(-0.25,0)
\linethickness{1pt}
\put(0.50,0.00){\circle*{.2}}
\put(0.50,0.50){\circle{1}}

\put(2,0.00){\circle*{.2}}
\put(2,0.50){\circle{1}}

\put(3.50,0.00){\circle*{.2}}
\put(3.50,0.50){\circle{1}}
\end{picture}
}

\newcommand{\graphgrra}{
\setlength{\unitlength}{1.8ex}
\begin{picture}(2,1)(-0.25,-0.25)
\linethickness{1pt}
\put(0.5,0){\circle*{.2}}
\put(1.5,0){\circle*{.2}}
\put(0.5,-0.5){\circle{1}}
\put(0.5,0.5){\circle{1}}

\linethickness{0.02em}
\put(0.5,0){\line(1,0){1}}
\end{picture}
}

\newcommand{\graphgrrb}{
\setlength{\unitlength}{1.8ex}
\begin{picture}(3.5,.75)(-0.25,-0.25)
\linethickness{1pt}
\put(1,0){\circle*{.2}}
\put(2,0){\circle*{.2}}
\put(0.5,0){\circle{1}}
\put(2.5,0){\circle{1}}

\linethickness{.5pt}
\put(1,0){\line(1,0){1}}
\end{picture}
}

\newcommand{\graphgrrc}{
\setlength{\unitlength}{1.8ex}
\begin{picture}(4,1)(-0.25,-0.25)
\linethickness{1pt}
\put(1,0){\circle*{.2}}
\put(2,0){\circle*{.2}}
\put(3,0){\circle*{.2}}
\put(0.5,0){\circle{1}}
\put(3,0.5){\circle{1}}

\linethickness{.5pt}
\put(1,0){\line(1,0){1}}
\end{picture}
}

\newcommand{\graphgrrd}{
\setlength{\unitlength}{1.8ex}
\begin{picture}(2.5,1)(-0.25,-0.25)
\linethickness{1pt}
\put(1,0.5){\circle*{.2}}
\put(0.5,-0.5){\circle*{.2}}
\put(1.5,-0.5){\circle*{.2}}
\put(0.5,0.5){\circle{1}}
\put(1.5,0.5){\circle{1}}

\linethickness{.5pt}
\put(0.5,-0.5){\line(1,0){1}}
\end{picture}
}

\newcommand{\graphgrre}{
\setlength{\unitlength}{1.8ex}
\begin{picture}(3,1)(-0.25,-0.25)
\linethickness{1pt}
\put(0.5,0){\circle*{.2}}
\put(2,0){\circle*{.2}}
\put(0.75,-0.5){\circle*{.2}}
\put(1.75,-0.5){\circle*{.2}}
\put(0.5,0.5){\circle{1}}
\put(2,0.5){\circle{1}}

\linethickness{.5pt}
\put(0.75,-0.5){\line(1,0){1}}
\end{picture}
}

\newcommand{\graphggra}{
\setlength{\unitlength}{1.8ex}
\begin{picture}(2.5,.75)(-0.25,-0.25)
\linethickness{1pt}
\put(0,0){\circle*{.2}}
\put(1,0){\circle*{.2}}
\put(0.5,0){\circle{1}}
\put(1.5,0){\circle{1}}
\end{picture}
}

\newcommand{\graphggrb}{
\setlength{\unitlength}{1.8ex}
\begin{picture}(3,.75)(-0.25,-0.25)
\linethickness{1pt}
\put(0,0){\circle*{.2}}
\put(1,0){\circle*{.2}}
\put(2,-0.5){\circle*{.2}}
\put(0.5,0){\circle{1}}
\put(2,0){\circle{1}}
\end{picture}
}

\newcommand{\graphggrc}{
\setlength{\unitlength}{1.8ex}
\begin{picture}(2.5,.75)(-0.25,-0.25)
\linethickness{.5pt}
\put(1,-0.25){\line(1,0){1}}
\put(1,-0.25){\line(-1,0){1}}

\put(1,-0.25){\circle*{.2}}
\put(0,-0.25){\circle*{.2}}
\put(2,-0.25){\circle*{.2}}
\linethickness{1pt}
\put(1,0.25){\circle{1}}
\end{picture}
}

\newcommand{\graphggrd}{
\setlength{\unitlength}{1.8ex}
\begin{picture}(2.5,.75)(-0.25,-0.25)
\linethickness{.5pt}
\put(0,0.75){\circle*{.2}}
\put(0,0.75){\line(0,-1){1}}
\put(0,-0.25){\circle*{.2}}
\put(0,-0.25){\line(1,0){1}}
\put(1,-0.25){\circle*{.2}}
\linethickness{1pt}
\put(1.5,-0.25){\circle{1}}
\end{picture}
}

\newcommand{\graphggre}{
\setlength{\unitlength}{1.8ex}
\begin{picture}(3,.75)(-0.25,-0.25)
\linethickness{.5pt}
\put(0,0.5){\circle*{.2}}
\put(0,0.5){\line(0,-1){1}}
\put(0,-0.5){\circle*{.2}}
\put(0,-0.5){\line(1,0){1}}
\put(1,-0.5){\circle*{.2}}
\put(2,-0.5){\circle*{.2}}
\linethickness{1pt}
\put(2,0){\circle{1}}
\end{picture}
}

\newcommand{\graphggrf}{
\setlength{\unitlength}{1.8ex}
\begin{picture}(2.5,1.25)(-0.25,-0.25)
\linethickness{.5pt}
\put(0,0.5){\line(1,0){1}}
\put(0,-0.5){\line(1,0){1}}
\linethickness{.5pt}
\put(0,0.5){\circle*{0.2}}
\put(0,-0.5){\circle*{0.2}}
\put(1,0.5){\circle*{0.2}}
\put(1,-0.5){\circle*{0.2}}
\put(1.5,0.5){\circle{1}}
\end{picture}
}

\newcommand{\graphggrg}{
\setlength{\unitlength}{1.8ex}
\begin{picture}(3,.75)(-0.25,-0.25)
\linethickness{.5pt}
\put(0,0.5){\line(1,0){1}}
\put(0,-0.5){\line(1,0){1}}
\linethickness{.5pt}
\put(0,0.5){\circle*{0.2}}
\put(0,-0.5){\circle*{0.2}}
\put(1,0.5){\circle*{0.2}}
\put(1,-0.5){\circle*{0.2}}
\put(1.5,0){\circle*{0.2}}
\put(2,0){\circle{1}}
\end{picture}
}

\newcommand{\graphggga}{
\setlength{\unitlength}{1.8ex}
\begin{picture}(1.5,1)(-0.25,-0.25)
\linethickness{.5pt}
\put(0,0){\circle*{.2}}
\put(0,0){\line(1,0){1}}
\put(1,0){\circle*{.2}}
\linethickness{1pt}
\put(0.5,0){\circle{1}}
\end{picture}
}

\newcommand{\graphgggb}{
\setlength{\unitlength}{1em}
\begin{picture}(1.5,.75)(-0.25,-0.25)
\linethickness{.5pt}
\put(0,-0.5){\circle*{.2}}
\put(0,-0.5){\line(1,1){1.0142}}
\put(0,-0.5){\line(1,0){1}}
\put(1,-.5){\circle*{.2}}
\put(1,-0.5){\line(0,1){1}}
\put(1,0.5){\circle*{.2}}
\end{picture}
}

\newcommand{\graphgggc}{
\setlength{\unitlength}{1.8ex}
\begin{picture}(2.5,.5)(-0.25,-0.25)
\linethickness{.5pt}
\put(0,0){\circle*{.2}}
\put(1,0){\circle*{.2}}
\put(1,0){\line(1,0){1}}
\put(2,0){\circle*{.2}}
\linethickness{1pt}
\put(0.5,0){\circle{1}}
\end{picture}
}

\newcommand{\graphgggd}{
\setlength{\unitlength}{1.8ex}
\begin{picture}(2.5,.75)(-0.25,-0.25)
\linethickness{.5pt}
\put(0,-0.5){\circle*{.2}}
\put(0,-0.5){\line(1,0){1}}
\put(1,-0.5){\circle*{.2}}
\put(1,-0.5){\line(1,0){1}}
\put(1,-0.5){\line(0,1){1}}
\put(1,0.5){\circle*{.2}}
\put(2,-0.5){\circle*{.2}}
\end{picture}
}

\newcommand{\graphggge}{
\setlength{\unitlength}{1.8ex}
\begin{picture}(1.5,1)(-0.25,-0.25)
\linethickness{.5pt}
\put(0,0.5){\circle*{.2}}
\put(0,0.5){\line(0,-1){1}}
\put(0,-0.5){\circle*{.2}}
\put(0,-0.5){\line(1,0){1}}
\put(1,-0.5){\circle*{.2}}
\put(1,-0.5){\line(0,1){1}}
\put(1,0.5){\circle*{.2}}
\end{picture}
}

\newcommand{\graphgggf}{
\setlength{\unitlength}{1.8ex}
\begin{picture}(1.5,1)(-0.25,-0.25)
\linethickness{.5pt}
\put(0,0.5){\circle*{.2}}
\put(0,-0.5){\circle*{.2}}
\put(1,0.5){\circle*{.2}}
\put(0,-0.5){\line(1,0){1}}
\put(1,-0.5){\circle*{.2}}
\linethickness{1pt}
\put(0.5,0.5){\circle{1}}\end{picture}
}

\newcommand{\graphgggg}{
\setlength{\unitlength}{1.8ex}
\begin{picture}(1.5,1)(-0.25,-0.25)
\linethickness{.5pt}
\put(0,0.75){\circle*{.2}}
\put(0,0.75){\line(1,0){1}}
\put(1,0.75){\circle*{.2}}
\put(1,0.75){\line(0,-1){1}}
\put(1,-0.25){\circle*{.2}}

\put(0,-0.75){\circle*{.2}}
\put(0,-0.75){\line(1,0){1}}
\put(1,-0.75){\circle*{.2}}
\end{picture}
}

\newcommand{\graphgggh}{
\setlength{\unitlength}{1.8ex}
\begin{picture}(1.5,1)(-0.25,-0.25)
\linethickness{.5pt}
\put(0,0.5){\circle*{.2}}
\put(0,0.5){\line(1,0){1}}
\put(1,0.5){\circle*{.2}}

\put(0,0){\circle*{.2}}
\put(0,0){\line(1,0){1}}
\put(1,0){\circle*{.2}}

\put(0,-0.5){\circle*{.2}}
\put(0,-0.5){\line(1,0){1}}
\put(1,-0.5){\circle*{.2}}
\end{picture}
}

\newcommand{\stacklr }[5]{ \,{}^{(#1)}_{\ #2}#3^{#4}_{#5}\,}

\newcommand{\Rmn}[1]{\uppercase\expandafter{\romannumeral #1}}

\allowdisplaybreaks
\makeindex

\begin{document}
\title{Exploring the transition from BCS to unitarity without Cooper pairs: the 
Pauli principle, normal modes and superfluidity. 
}
\author{D. K. Watson \\
University of Oklahoma \\ Homer L. Dodge Department of Physics and Astronomy \\
Norman, OK 73019}
\date{\today}

\begin{abstract}
The transition from the weakly interacting BCS regime to the strongly
interacting unitary regime is explored for ultracold 
trapped Fermi gases assuming a 
normal mode description of the gas instead of the conventional Cooper pairing.
The Pauli principle is applied ``on paper'' by using specific normal
mode assignments. 
Energies, entropies, critical temperatures, and an excitation
frequency are studied 
and compared to existing results in the literature. These normal modes have been
derived analytically for $N$ identical, confined particles 
from a first-order $L=0$ group theoretic solution of a
 three-dimensional Hamiltonian with a general two-body 
interaction. In previous studies, normal modes were able to describe 
the unitary regime obtaining ground state energies comparable to benchmark 
results and thermodynamics quantities in excellent agreement with experiment.
In a recent study, the behavior of the normal mode frequencies was 
investigated for Hamiltonians with a range
of interparticle interaction strengths from BCS to unitarity in the first test
of this approach beyond the unitary regime, and a
microscopic basis of the large excitation gaps and universal
behavior at unitarity was proposed. Based on the success of these earlier
studies, the 
current paper continues to explore the ability of normal modes to
describe superfluidity along the 
BCS to unitarity transition.   
The results confirm earlier conclusions that the physics of superfluidity
can be described using normal modes 
across a wide range of interparticle interaction strengths and offers
 an alternative to the two-body pairing models commonly used
to describe superfluidity along this transition.

\end{abstract}


\maketitle

\section{Introduction}

The BCS to unitarity transition for
ultracold gaseous fermions has been investigated 
both experimentally and theoretically 
since this transition was first
achieved in the laboratory\cite{jin1,jin2,zwierlein1,zwierlein2,grimm1,salomon1,thomas1,hulet1,jochim1}. Theoretical methods typically
assume that the atomic fermions are forming Cooper pairs 
to explain the emergence of superfluid 
behavior\cite{giorgini1,randeria1,leggett1,leggett2,bcs,eagles,nozieres}.
When a Feshbach resonance is tuned to weak interactions, the neutral atoms
 bind into loosely-bound pairs whose
size decreases as the interparticle interaction strength increases 
toward the unitarity. Eventually 
diatomic molecules are produced that condense in the BEC regime.
In materials that support superconductivity, the binding of electrons
into Cooper pairs at long distances is thought to be 
mediated by phonon interactions
in the underlying material producing a weak 
attraction\cite{bcs,leggett1,leggett2,eagles,nozieres}.

In a series of recent papers, the ability of normal modes to describe 
superfluidity  has been investigated for ultracold Fermi 
gases\cite{prl,emergence,prafreq}. Initial
studies were in the unitary regime where results for ground state energies 
comparable to benchmark results\cite{prl} and thermodynamic quantities in 
excellent
agreement with experiment were obtained\cite{emergence}. 
The $N$-body analytic frequencies of the normal modes 
were studied from the BCS regime to unitarity\cite{prafreq}
revealing the emergence of excitation gaps that increased from extremely
small gaps deep in the BCS regime to a maximum at unitarity as observed
in experiments. The
microscopic dynamics responsible for the emergence of these gaps was
investigated using the analytic forms of the normal mode functions 
and a 
microscopic basis for the universal behavior at unitarity was proposed.
These calculations have modest numerical requirements since the 
$N$-body normal
modes have been derived in analytic form using group theoretic 
techniques\cite{paperI,JMPpaper}.

In this paper, I continue to explore the ability of normal modes to
describe superfluidity from BCS to unitarity 
by determining various properties along this transition
including ground state energies, thermodynamic entropies, 
critical temperatures, and the breathing excitation frequency, 
comparing to both experiment
 and theory. 
This approach models the physics by assuming 
many-body pairing exhibited through normal modes, 
i.e. coherent, collisionless
motion of the fermions that minimizes interparticle interactions and
makes two-body pairing irrelevant since it
is impossible to discern which fermion is
paired with another fermion.

 Normal mode functions naturally
provide simple, coherent macroscopic wave functions that maintain
 phase coherence over the entire ensemble, and give rise to
 ``quasiparticles'' defined by
the excitations between the modes.
At ultracold temperatures, only the lowest two types of normal
mode frequencies are relevant,  gapless phonon modes with 
very low frequencies and  particle-hole excitation modes.

Normal mode motions exist at all scales in our universe from  
vibrating crystals\cite{NM4} to  oscillating black holes\cite{NM10}. 
The particles in a normal mode move in synchrony with the same 
frequency and 
phase, allowing a description of the complex, simultaneous motions of many 
interacting particles in terms of collective behavior.  
These modes are a manifestation of 
the widespread appearance 
of vibrational motions that occur in nature in diverse media and across many 
orders of magnitude\cite{NM1,NM2,NM3,NM4,NM5,NM6,NM7,NM8,NM9,NM10,
NM11,NM12,NM13}. When higher order effects are small,
vibrational behavior couples into stable collective motion, thus 
incorporating the many-body effects of large ensembles into simple dynamic 
motions. 
These collective motions correspond to the eigenfunctions of an  
approximate Hamiltonian  and thus will possess some stability 
over time. Normal modes will 
reflect the symmetry that is present in this approximate Hamiltonian
and can offer beyond-mean-field analytic many-body
solutions  
and physical intuition into the microscopic dynamics responsible for diverse
phenomena.

\subparagraph{SPT formalism.} The formalism used to obtain these normal modes
 is called symmetry-invariant
perturbation theory (SPT), a first-principle, non-numerical method which
use group theoretic and graphical techniques to solve many-body problems.
 This perturbation formalism was initially 
developed for   
bosons\cite{FGpaper,energy,paperI,JMPpaper,laingdensity,test,toth} and
 has been formulated through first order for $L=0$, 
three-dimensional systems with completely general interaction potentials
and spherically-symmetric confining potentials. 
Unlike conventional methods for which the resources for 
an exact solution of the quantum $N$-body wave function
 scale exponentially with $N$,
typically doubling for every 
particle added\cite{liu,montina2008}, the SPT
approach employs symmetry to attack the $N$-scaling
problem\cite{test}. This is accomplished by formulating a 
perturbation series about a large-dimension
configuration whose point group is isomorphic to the symmetric group $S_N$\,, 
and then evaluating the series for $D=3$.
Group theoretic techniques are used to extract the part of the problem
that scales exponentially with 
complexity as a pure mathematical problem (cf. the Wigner-Eckart theorem),
which can then be solved as a function of
$N$, rendering the interaction dynamics containing the ``physics''
independent of $N$\cite{rearrangeprl,complexity}. 
The perturbation terms are evaluated for
large dimension where the structure has maximum 
symmetry yielding terms that are invariant 
under the $N!$ operations of the $S_N$ point group.
This strategy produces a problem at each order 
that, in principle, can be solved exactly, analytically using symmetry. 
Although extremely challenging, the mathematical work 
at each order can be saved and used
to study a problem with a new interaction potential 
significantly reducing numerical demands.

The symmetry constraints are enforced by using a tensor basis 
that is  small, complete and $S_N$-invariant.
(A proof of the completeness is in Ref~\cite{test}.)
 As $N$ increases, the $N!$ operations of $S_N$
place increasing constraints on the basis whose size, therefore,
 does not grow with $N$.
First order needs only seven elements while the next order requires
 twenty-five elements regardless of the $N$ value. 
(See Section ~\ref{subsec:binary}.)
 These tensor basis functions are called
``binary invariants'' reflecting their invariance under the symmetric group
operations and the use of $1's$ and $0's$ in their tensor definition.
When this basis is used for the perturbation expansion, the
Hamiltonian will automatically be invariant at each 
order.

The SPT formalism accounts for every two-particle interaction rather
than assuming an average interaction and can be applied
 to strongly interacting systems 
 since the perturbation is not dependent on the 
interaction. Even at the lowest perturbation order, the SPT method includes
beyond-mean-field effects underlying the excellent
results achieved at first order using this SPT
method\cite{energy,prl,emergence} as well as 
earlier dimensional approaches\cite{herschbach1,herschbach2,loeser,kais1,kais2}.

This formalism has also been implemented for a
 model problem of harmonically-confined, harmonically-interacting 
particles that is exactly solvable\cite{test,toth,harmoniumpra,partition}. 
Agreement to ten or more digits of accuracy was found for the
 wave function compared to the exact  
wave function obtained independently, verifying this general many-body 
formalism 
for a three-dimensional, many-body system that is fully-interacting  
\cite{test} including the formulas derived analytically
for the $N$-body normal mode coordinates and frequencies.

\subparagraph{Application to fermions.} The application of SPT to fermions has been developed in the last 
seven years\cite{prl,harmoniumpra,partition,emergence,annphys,prafreq}. Initially,
these studies focused on the unitary regime where I obtained excellent
values for the ground state energies comparable to 
numerically intensive benchmark methods\cite{prl} 
as well as 
thermodynamic quantities for
the energy, entropy, and specific heat in close agreement with 
experiment\cite{emergence}. The lambda transition in the specific heat was clearly seen 
defining a critical temperature for the
onset of superfluidity in the unitary regime.  
These results support the validity of this normal mode 
description of superfluidity and 
the role of the Pauli principle in low temperature dynamics\cite{emergence}.
The heavy numerical demands of enforcing antisymmetry 
in fermion systems in conventional theoretical approaches
is avoided in the SPT approach by enforcing the Pauli principle
 ``on paper'' using specific occupations of the
normal modes   
at first order\cite{prl,harmoniumpra,emergence,partition}. 
(See Section~\ref{subsec:pauli}.)  Beyond-mean-field ground\cite{prl}
 and excited state\cite{emergence}
energies and their degeneracies have been calculated 
 enabling the determination of a partition 
function and the calculation
 of thermodynamic 
quantities\cite{partition,emergence}. An accurate partition function 
requires many states chosen  
from the 
infinite spectrum by the
Pauli principle,
 thus relating the Pauli principle to 
many-body interaction dynamics through the normal modes.

\subparagraph{The physical character of the normal modes.} 
The close agreement with experiment for energies and thermodynamic
observables in the unitary regime motivated an investigation into the
physical character of the normal modes with the goal of obtaining
 insight into the dynamics of cooperative motion\cite{annphys} and the 
universal behavior at unitarity.
In a recent paper, I used the analytic $N$-body 
normal mode coordinates to study the character of the five types of 
normal modes\cite{annphys}, investigating their evolving
 motions as a function of $N$,
from small $N$ to large ensembles. I analyzed the contributions of the 
particles individually to the collective motion, making some general 
observations
based on symmetry 
considerations, and then focusing on the unitary
 Hamiltonian. 

This study found a smooth evolution  as $N$ increases from the expected
behavior for few-body systems 
whose motions are analogous to
those of molecular equivalents such as ammonia and 
methane,  to the coherent motions observed in 
large $N$ ensembles.
Furthermore, the transition from 
few-body
to large $N$ behavior happens at surprisingly low values of 
$N$ ($N \approx 10$) validating the results of numerous few-body
studies\cite{adhikari,hu6,hu7,hu8,blume1,blume2,levinsen,grining}. 
This evolution in character from few-body to large ensembles
is dictated by rather simple analytic forms (See Ref.~\cite{emergence} after
Eq. (31).)
that nevertheless take into account
the complicated interplay of the particles as they interact and
cooperate to create coherent
macroscopic motion. 
This evolution of behavior was dependent primarily on the 
symmetry present in the Hamiltonian, and thus could be relevant for 
diverse
phenomena at different scales if the same symmetry exists or is dominant. 
Also, two phenomena were found 
that could support the emergence and 
stabilization of 
collective behavior for the unitary regime.

\subparagraph{The evolution of the normal mode frequencies.}
In my most recent study, I extended my investigation 
beyond unitarity, studying the evolution of the 
analytic frequencies as a function of the interparticle interaction 
strength,
$\bar{V}_{\mathtt{int}}\sim \bar{V}_0$, (See Eq.~(\ref{eq:int}).)
for fixed ensemble sizes
from BCS to unitarity\cite{prafreq}.
 I focused on larger values of $N$ relevant to laboratory studies 
of this transition that use a Feshbach resonance to tune the interaction.  
This work offers
 insight into the microscopic behavior leading to large
gaps and universal behavior at unitarity and offers
the possibility of controlling
the appearance and stability of excitation gaps by fine tuning 
system parameters.

This analysis revealed the appearance of excitation gaps  that increase
as $\bar{V}_0$ increases. At the independent particle limit where interactions
vanish,
all five frequencies limit to the same value at twice the trap
frequency.
As $\bar{V}_0$ turns on, the frequencies slowly spread out 
producing 
 gaps that are maximal at unitarity.  

\subparagraph{The microscopic dynamics underpinning universal behavior.}
The microscopic dynamics underpinning
the emergence
of universal behavior at unitarity and the convergence of the 
angular frequencies to integer multiples of the trap frequency
were investigated using two approaches. First,
the effect of different Hamiltonian terms on the analytic
frequencies was tracked as $\bar{V}_0$ changes. Second, 
the motion of each particle
in the normal mode coordinate was analyzed to discern
how the ensemble rearranges microscopically as interactions
begin and cooperative behavior emerges. At unitarity, the 
angular particle-hole frequency converges to the frequency of the trap and 
sets up an
 evenly-spaced spectrum identical to the independent particle
spectrum. Also, as $\bar{V}_0$ increases, correlation of the particles
increases minimizing the interparticle interactions.
These results are consistent with the universal behavior expected at unitarity
and offer insight into its microscopic basis.

The study of the frequencies across the BCS to unitarity transition
suggested that normal modes are able to describe
the physics of ultracold Fermi gases including 
superfluidity for a range of interaction strengths
 and to offer insight into the underlying microscopic basis 
 without the assumption
of two-body pairing.  
In the current paper, I now look more closely at the ability of these 
first-order normal mode
solutions to accurately describe the BCS to unitarity transition 
by determining various properties along this transition. 
The value of $\bar{V}_0$ is scaled so
$\bar{V}_0 = 1.0$ in the unitary regime, while 
the BCS region is loosely-defined by very weak interparticle interactions, 
e.g. $\bar{V}_0 \approx 10^{-6}$. 
This potential is defined in Section~\ref{sec:SPT}
with a more detailed description in Appendix A in Ref.~\cite{annphys}.

\section{The General N-body problem: A Group Theoretic and 
Graphical Approach }\label{sec:SPT}

This section contains a summary of the SPT formalism including
the symmetry coordinates, the normal modes 
and their frequencies
presented in Refs.~\cite{paperI,FGpaper}.

\subsection{The Hamiltonian}

For $N$ interacting particles, the Schr\"odinger equation in 
$D$ dimensions is:
\begin{equation} \label{eq:generalH} 
H \Psi  =  \left[ \sum\limits_{i=1}^{N} h_{i} +
\sum_{i=1}^{N-1}\sum\limits_{j=i+1}^{N} g_{ij} \right] \Psi = E
\Psi \,,  
\end{equation} 
\begin{equation} \label{eq:generalH1} 
\begin{array}{rcl}
h_{i} & = & -\frac{\hbar^2}{2
m_{i}}\sum\limits_{\nu=1}^{D}\frac{\partial^2}{\partial
x_{i\nu}^2} +
V_{\mathtt{conf}}\left(\sqrt{\sum\nolimits_{\nu=1}^{D}x_{i\nu}^2}\right)
\,,  \\
g_{ij} & = & V_{\mathtt{int}}\left(\sqrt{\sum\nolimits_{\nu=1}^{D}\left(x_{i\nu}-x_{j\nu}
\right)^2}\right),  
\end{array}
\end{equation}
\noindent where $h_{i}$ is the single-particle Hamiltonian,
$g_{ij}$ a two-body interaction potential,
 $x_{i\nu}$ the $\nu^{th}$ Cartesian component
of the $i^{th}$ particle, and 
$V_{\mathtt{conf}}$ is a spherically-symmetric confining 
potential\cite{paperI,FGpaper,JMPpaper}. 
Defining internal coordinates as the $D$-dimensional scalar radii $r_i$ 
of the $N$
particles from the center of the trap and the
cosines $\gamma_{ij}$ of the $N(N-1)/2$ interparticle angles between the radial
vectors:

\begin{equation}\label{eq:int_coords}
\renewcommand{\arraystretch}{1.5}
\begin{array}{rcl} r_i & = &\sqrt{\sum_{\nu=1}^{D} x_{i\nu}^2}\,, \;\;\; (1 \le i \le
N)\,,
\;\;\; 
\\ \gamma_{ij} & = & cos(\theta_{ij})=\left(\sum_{\nu=1}^{D}
x_{i\nu}x_{j\nu}\right) / r_i r_j\,,
\end{array}
\renewcommand{\arraystretch}{1}
\end{equation}
\noindent $(1 \le i < j \le N)$\,, the  Schr\"odinger equation is transformed 
from Cartesian to internal coordinates.

A similarity transformation\cite{avery} removes the
 first-order derivatives, and a scale factor, $\kappa(D)= D^2 \bar{a}_{ho}$, 
is used to regularize the large-dimension limit of the Schr\"odinger equation 
by defining dimensionally-scaled oscillator length units   
where $\bar{a}_{\mathtt{ho}}=\sqrt{\frac{\hbar}{m\bar{\omega}_{\mathtt{ho}}}}$ and
${\bar{\omega}_{\mathtt{ho}}}=D^3{\omega_{\mathtt{ho}}}$.
Substituting the scaled variables, $\bar{r}_i = r_i/\kappa(D)$, 
with $\bar{E} = \frac{E}{\hbar\bar{\omega}_{\mathtt{ho}}}$ and 
$\bar{H} = \frac{H}{\hbar\bar{\omega}_{\mathtt{ho}}}$, 
into the similarity-transformed Schr\"odinger equation gives:
\begin{equation} \label{eq:scaleH_BEC}
\bar{H} \Phi =
\left(\delta^2\bar{T}+\bar{U}+\bar{V}_{\mathtt{conf}}+\bar{V}_{\mathtt{int}}\right)\Phi = \bar{E}\, \Phi\,.
\end{equation}
where 
\begin{equation}\label{eq:Tbar_BEC}
\begin{split}
\bar{T} &  = \sum\limits_{i=1}^{N} \Bigl( -\frac{1}{2}\frac{\partial^2}
{{\partial \bar{r}_i}^2} \\
& - \frac{1}{2 \bar{r}_i^2}\sum\limits_{j\not=i}\sum\limits_{k\not=i}
\frac{\partial}{\partial\gamma_{ij}}(\gamma_{jk}-\gamma_{ij}\gamma_{ik})
\frac{\partial}{\partial\gamma_{ik}} \Bigr) \,,
\end{split}
\end{equation}
\begin{eqnarray}
\bar{U}&=&\sum\limits_{i=1}^{N}\left(\frac{\delta^2N(N-2)+(1-\delta(N+1))^2 \left(\frac{\Gamma^{(i)}}{\Gamma}\right)}{8 \bar{r}_i^2}\right) \,, \label{eq:Ubar_BEC}
\\
\bar{V}_{\mathtt{conf}}&=&\sum\limits_{i=1}^{N}\frac{1}{2}\bar{r}_i^2
 \label{eq:confC}
\\
\bar{V}_{\mathtt{int}}&=&  \frac{\bar{V}_{0}}{1-3b'\delta}
\sum\limits_{i=1}^{N-1}\sum\limits_{j=i+1}^{N}
\left(1-\tanh\Theta_{i,j} \right) \,, \ \label{eq:int}
\end{eqnarray}
\noindent and $\delta=1/D$, $\Gamma$ is the Gramian determinant which has 
elements $\gamma_{ij}$ (See
Appendix D in Ref~\cite{FGpaper}.), $\Gamma^{(i)}$ the determinant which has
 the  $i^{th}$ row
and column deleted, and $\hbar=m=1$.  
The barred quantities are scaled by $\kappa(D)$.

The interaction potential, $\bar{V}_{\mathtt{int}}$, reduces
 to a square well for $D=3$.
The value of the constant $b'$ yields a 
scattering length of infinity when $\bar{V}_0 = 1.0$.  
$\bar{V}_0$ is scaled to smaller
values to reach the weaker 
interactions of the BCS regime. The argument $\Theta_{ij}$ is:
\begin{equation}\label{eq:thetaij_scaled}
\Theta_{ij}=\frac{\bar{c}_0}{1-3\delta}
\left(\frac{\bar{r}_{ij}}{\sqrt{2}}-\bar{\alpha}-3\delta\left(\bar{R}-\bar{\alpha} \right) \right)
\,,
\end{equation}
where 
$\bar{r}_{ij}={\sqrt{{\bar{r}_i}^2+
{\bar{r}_j}^2-2\bar{r}_i\bar{r}_j\gamma_{ij}}}\,$
is the interatomic separation,
$\bar{R}$ is the  dimensionally-scaled range of the square-well potential, 
and $\bar{\alpha}$ is a 
constant that softens the
potential as $D \rightarrow \infty$.
$R$ is chosen so 
$R<<a_{ho}$ ($a_{ho}=\sqrt{\hbar/(m\omega_{ho})}$) and is
 extrapolated 
to zero-range interaction.

At the $D\to\infty$ limit, the second derivative terms of the kinetic
energy drop out
 resulting in a static zeroth-order
problem with an effective potential, $\bar{V}_{\mathtt{eff}}$:
\begin{eqnarray}\label{eq:veff_BEC}
\bar{V}_{\mathtt{eff}}(\bar{r},\gamma;\delta)&=&
\sum\limits_{i=1}^{N}\left(\bar{U}(\bar{r}_i;\delta)
+\bar{V}_{\mathtt{conf}}(\bar{r}_i;\delta)\right) \nonumber \\
&& +\sum\limits_{i=1}^{N-1}\sum\limits_{j=i+1}^{N}
\bar{V}_{\mathtt{int}}(\bar{r}_i,\gamma_{ij};\delta)\,.
\end{eqnarray}
The minimum of  $\bar{V}_{\mathtt{eff}}$ corresponds to a large-dimension 
maximally-symmetric configuration that has all radii, $\bar{r}_i$, and 
angle  cosines, $\gamma_{ij}$, of the particles equal, i.e. when $D\to\infty$, 
$\bar{r}_{i}=\bar{r}_{\infty} \;\; (1 \le i \le N)$ and
$\gamma_{ij}={\gamma}_\infty \;\; (1 \le i < j \le N)$.
These parameters are solved using 
two  minimum conditions: $\left(\frac{\partial \bar{V}_\text{eff}}
{\partial\bar{r}_{i}}\right)\Biggr|_{\infty}=0, \,\,
\left(\frac{\partial \bar{V}_\text{eff}}
{\partial\gamma_{ij}}\right)\Biggr|_{\infty}= 0\,.$
\noindent Using the definition of $\bar{V}_\textrm{eff}$, 
two equations in $\bar{r}_\infty$ and $\gamma_\infty$ yield:
$ \bar{r}_\infty=\frac{1}{\sqrt{2} \sqrt{1+(N-1) \gamma _{\infty }}}\,,$
while $\gamma_{\infty}$ is solved using 
the transcendental equation: 

\begin{equation}\label{eq:gammaeq0bec}
   \frac{\gamma _{\infty } \left(2+(N-2) \gamma _{\infty
   }\right)}{\left(1-\gamma _{\infty }\right){}^{3/2}
   \sqrt{1+(N-1) \gamma _{\infty }}}
   +
   \bar{V}_0\,\text{sech}^2\left(\Theta _{\infty}\right)\Theta _{\infty }'
   =
   0\,.
\end{equation}
\noindent In the large-$D$ limit ($\delta \rightarrow 0$), 
the argument $\Theta_{ij}$ becomes: $\Theta_\infty
=\Theta_{ij}\Biggr|_{\infty}
=
\bar{c}_0 \left(\sqrt{1-\gamma _{\infty }}
   \,\bar{r}_{\infty }-\bar{\alpha}\right) \,.$

\subsection{The Dimensional Expansion}

The energy minimum as $\delta \rightarrow 0$, $\bar{E}_\infty$,
is the 
starting 
point for the $1/D$ expansion.
The $N(N+1)/2$ internal coordinates, $\bar{r}_{i}$ and  $\gamma_{ij}$,
 are expanded as: 
$\bar{r}_{i} = \bar{r}_{\infty}+\delta^{1/2}\bar{r}'_{i}$
and $\gamma_{ij} =
{\gamma}_{\infty}+\delta^{1/2}{\gamma}'_{ij}$ setting up a
power series in $\delta^{1/2}$ about
the $D\to\infty$ symmetric minimum.
The primed variables, $\bar{r}'_{i} $ and $\overline{\gamma}'_{ij}$\,, are 
dimensionally-scaled internal {\it displacement} coordinates:
\noindent The expansions of the Hamiltonian, wave function,
and energy in powers of $\delta^{1/2}$ are:
\begin{equation} \label{eq:dpt_exp}
\renewcommand{\arraystretch}{2} \begin{array}{r@{}l@{}c}
{\displaystyle \bar{H} = \bar{H}_{\infty} + \delta^{\frac{1}{2}}
\, \bar{H}_{-1} + \delta} & {\displaystyle \, \sum_{j=0}^\infty
\left(\delta^{\frac{1}{2}}\right)^j 
\bar{H}_j } & \\
{\displaystyle \Phi(\bar{r}_i,\gamma_{ij}) = } & {\displaystyle \,
\sum_{j=0}^\infty \left(\delta^{\frac{1}{2}}\right)^j 
\Phi_j }& \\
{\displaystyle \bar{E} = \bar{E}_{\infty} + \delta^{\frac{1}{2}}
\, \bar{E}_{-1} + \delta} & {\displaystyle \, \sum_{j=0}^\infty
\left(\delta^{\frac{1}{2}}\right)^j 
\bar{E}_j } &
 \,,
\end{array}
\renewcommand{\arraystretch}{1}
\end{equation}
where
\begin{eqnarray}
 \bar{H}_{\infty} & = & \bar{E}_{\infty}  \\
\bar{H}_{-1} & = &  \bar{E}_{2n-1} =  0\,, \label{eq:mone_H} \\
\bar{H}_{0} & = & -\frac{1}{2}\stacklr{0}{2}{G}{}{\nu_1,\nu_2}
\partial_{{\bar{y}^\prime}_{\nu_1}}
\partial_{{\bar{y}^\prime}_{\nu_2 }} +
\frac{1}{2} 
\stacklr{0}{2}{F}{}{\nu_1,\nu_2}
\bar{y}^\prime_{\nu_1 } \label{eq:harm_H}
\bar{y}^\prime_{\nu_2}\nonumber \\  
&& +\stacklr{0}{0}{F}{}{} \,, \label{eq:H0y} \\
\bar{H}_{1} & = &-\frac{1}{2}\stacklr{1}{3}{G}{}{\nu_1,\nu_2,\nu_3}
\bar{y}^\prime_{\nu_1 } \partial_{{\bar{y}^\prime}_{\nu_2 }}
\partial_{{\bar{y}^\prime}_{\nu_3 }}
-\frac{1}{2}\stacklr{1}{1}{G}{}{\nu}
\partial_{{\bar{y}^\prime}_\nu}\nonumber  \\ 
&&+\frac{1}{3!}\stacklr{1}{3}{F}{}{\nu_1,\nu_2,\nu_3}
\bar{y}^\prime_{\nu_1 } \bar{y}^\prime_{\nu_2}
\bar{y}^\prime_{\nu_3 } 
 +\stacklr{1}{1}{F}{}{\nu}
\bar{y}^\prime_\nu \,. \label{eq:one_H}
\end{eqnarray}

The superprescript in parentheses on the $\bm{F}$ and $\bm{G}$ tensors 
in Eqs.~(\ref{eq:H0y})-(\ref{eq:one_H}) 
denotes the order in $\delta^{1/2}$ in the sum over $j$ in 
Eq.~(\ref{eq:dpt_exp}).
 The subprescripts give the rank, $R$, of
the tensors. The $G$ elements are defined from the first-order 
derivative terms, ${\bar T}$, of the Hamiltonian  
while
the $F$ elements contain the first-order potential terms from 
${\bar V_{\mathtt{eff}}}$. 
Appendix B of Ref.~\cite{energy} gives formulas for the $F$ and $G$ 
elements.

\subsection{The Binary Invariant Basis} \label{subsec:binary}

A tensor basis of binary
invariants is used to obtain the $N$-body perturbation solutions  exactly 
 through first order\cite{JMPpaper,test,toth}. 
A rank $R$ tensor has dimension $[N(N+1)/2]^R$ 
and is comprised
of $[N(N+1)/2]^R$ elements, either 1's or 0's positioned so 
the tensor is invariant under the $N!$ operations of $S_N$. At each order,
the binary invariants constitute a complete basis  spanning the
tensor space. Each binary invariant
can be represented by an unlabelled multiloop graph (with no unattached
vertices)\cite{epaps}. 

For example, irrespective of the value of $N$, only seven graphs are required for the 
kinetic
energy terms for $\bar H_0$:

\begin{eqnarray}\label{eq:GXX}
\mathbb{G}_{rr}&=&\{\graphrra,\graphrrb\},\,\,\,
\mathbb{G}_{\gamma r}=\{\graphgra,\graphgrb\},\,\,\,\\
\mathbb{G}_{\gamma \gamma}
&=&\{\graphgga,\graphggb,\graphggc\}
\end{eqnarray}
\noindent while twenty five graphs are needed for 
$\bar H_1$:

\begin{eqnarray}\label{eq:GX}
\mathbb{G}_{r}&=&\{\graphr\},\,\,\,\,\,\,\,\,\,
\mathbb{G}_{\gamma}=\{\graphgamma\}
\end{eqnarray}
\begin{eqnarray}\label{eq:GXXX}
\mathbb{G}_{rrr}&=&\{\graphrrra,\graphrrrb,\graphrrrc\}
\\
\mathbb{G}_{\gamma rr}
&=&\{\graphgrra,\graphgrrb,\graphgrrc,\graphgrrd,\graphgrre\}
\nonumber\\
\mathbb{G}_{\gamma \gamma r}
&=&\{\graphggra,\graphggrb,\graphggrc,\graphggrd,\graphggre,\graphggrf,\graphggrg\}
\nonumber\\
\mathbb{G}_{\gamma \gamma \gamma}
&=&\{\graphggga,\graphgggb,\graphgggc,\graphgggd,\graphggge,\graphgggf,\graphgggg,\graphgggh\}\,.
\nonumber
\end{eqnarray}
The above graphs correspond to particular binary invariants and
are grouped according to the number of
loop edges ($r$) and  straight edges ($\gamma$). 
An EPAPS
document contains explicit expressions\cite{epaps}.

\subsection{Symmetry Coordinates and Normal Modes}
According to Eqs.~(\ref{eq:dpt_exp}) and ~(\ref{eq:harm_H}), 
$\bar{H}_{0}$ contains contributions
from the Hamiltonian through first order 
in the displacements from the maximally-symmetric structure. This expansion
thus includes first-order effects from all the Hamiltonian terms
including the
interparticle interaction. Since $\bar{H}_{0}$ has the form
of a multidimensional harmonic oscillator,
the first-order wave function can be expressed in terms of the 
 normal mode basis whose frequencies and coordinates include effects of the 
many-body interactions of the particles through first order.
Since $\bar{H}_{0}$ is invariant under
$S_N$, the normal modes transform under irreducible
representations (irreps.) of the $S_N$ group. 
For the $\bar{\bm{r}}'$ vector, the irreps. are 
 $[N]$ and
$[N-1, \hspace{1ex} 1]$\,, while for
the $\overline{\bm{\gamma}}'$ vector, the irreps. are $[N]$\,,
$[N-1, \hspace{1ex} 1]$\,, and $[N-2, \hspace{1ex} 2]$.

The normal mode coordinates and their frequencies are obtained using 
a quantum chemistry method, the FG method
developed by Wilson in 1941\cite{wilson}, 
 which has been used extensively   
to study molecular normal mode behavior\cite{dcw}. A review is presented in 
Appendix A of Ref.~\cite{FGpaper}.
The determination of the normal modes and their frequencies
 has been achieved analytically by using group theoretic techniques. 
The $N(N+1)/2$ roots, ${\bar{\omega}_{\mu}}^2$,
 are highly degenerate due to the $S_N$ symmetry,
producing five distinct roots  and five types of normal modes corresponding
 to five irreducible representations of
$S_N$\cite{hamermesh,WDC} labelled by  
${\bf 0^+, 0^-, 1^+, 1^-, 2}$\cite{FGpaper}. The $N(N-3)/2$  
normal modes of type ${\bf 2}$ are 
phonon modes; the $N-1$
   modes of type ${\bf 1^-}$ exhibit
single-particle i.e. particle-hole radial excitation behavior;  
the $N-1$  normal modes
of type ${\bf 1^+}$
have single-particle/particle-hole
angular excitation behavior; the single ${\bf 0^+}$  normal mode 
is a symmetric bend/center of mass motion,
and the single ${\bf 0^-}$ normal mode is a symmetric stretch/
breathing motion. These motions are analyzed in detail in
Ref.~\cite{annphys}. 
The energy through first order in $\delta = 1/D$ is:
\cite{FGpaper}
\begin{equation}
\overline{E} = \overline{E}_{\infty} + \delta \Biggl[
\sum_{\renewcommand{\arraystretch}{0}
\begin{array}[t]{r@{}l@{}c@{}l@{}l} \scriptstyle \mu = \{
  & \scriptstyle \bm{0}^\pm,\hspace{0.5ex}
  & \scriptstyle \bm{1}^\pm & , 
  &  \,\scriptstyle \bm{2}   \scriptstyle  \}
            \end{array}
            \renewcommand{\arraystretch}{1} }
(n_{\mu}+\frac{1}{2} d_{\mu})
\bar{\omega}_{\mu} \, + \, v_o \Biggr] \,, \label{eq:E1}
\end{equation}

\noindent where  $n_{\mu}$ is the total normal mode quanta
with frequency $\bar{\omega}_{\mu}$; 
 $\mu$ the normal mode label (${\bf 0^+, 0^-, 1^+, 1^-, 2}$),
and $v_o$  a constant.
The
multiplicities of the normal modes are:
$d_{{\bf 0}^+} = 1, \hspace{1ex} d_{{\bf 0}^-} = 1,\;
d_{{\bf 1}^+} = N-1,\;  d_{{\bf 1}^-} = N-1,\;
d_{{\bf 2}} = N(N-3)/2$.

The normal modes for the $\alpha=[N]$  and 
$[N-1, \hspace{1ex} 1] $ sectors in terms of symmetry coordinates 
 $[{\bm{S}}_{X'}^{\alpha}]_\xi$\, are:\cite{paperI}

\begin{equation} \label{eq:qnpfullexp}
{\bm{q}'}_\pm^\alpha = c_\pm^{\alpha} \left(
\cos{\theta^\alpha_\pm} \, [{\bm{S}}_{\bar{\bm{r}}'}^{\alpha}]_\xi
\, + \, \sin{\theta^\alpha_\pm} \,
[{\bm{S}}_{\overline{\bm{\gamma}}'}^{\alpha}]_\xi \right)
\end{equation}

\noindent where 
$\cos{\theta^\alpha_\pm}$ and  $\sin{\theta^\alpha_\pm}$
are mixing coefficients and
the $\pm$ 
refer to $0^+$ and $0^-$ for the $[N]$ sector and
$1^+$ and $1^-$ for the $[N-1,1]$ sector.
 The ${\bf 2}$ normal mode is:
\begin{equation} \label{eq:qnm2fullexp}
{\bm{q}'}^{[N-2, \hspace{1ex} 2]} = c^{[N-2, \hspace{1ex} 2]}
{\bm{S}}_{\overline{\bm{\gamma}}'}^{[N-2, \hspace{1ex} 2]} \,.
\end{equation}
The symmetry coordinates 
were derived in  Ref.~\cite{paperI} and are summarized after Eq. (31) in 
Ref.~\cite{emergence}.

The large degeneracies of the
frequencies reflect the very high degree of symmetry of
the $\bm{F}$\, and $\bm{G}$\, matrices whose
 elements are evaluated for the large-dimension,
maximally-symmetric configuration that has a single value for all radii 
$\bar{r}_\infty$ and all
angle cosines, ${\gamma}_{\infty}$. 
This yields invariant matrices 
under the $N!$ operations of particle exchanges 
effected by the symmetric group, $S_N$.

From Eq.~(\ref{eq:qnpfullexp}), the normal modes in both
the $[N]$ and $[N-1,1]$ sectors will have both radial and angular behavior 
 depending on the mixing angles.
The $[N-2,2]$ normal modes (Eq.~(\ref{eq:qnm2fullexp})) 
are purely angular since this sector has
no  $\bar{\bm{r}}'$ symmetry coordinates. 
The amount of mixing in a normal coordinate
 depends, of course, on the first-order Hamiltonian terms 
 and was investigated in previous studies\cite{annphys,prafreq}.

\subsection{The Pauli Priniple} \label{subsec:pauli}

The energy expression, Eq.~(\ref{eq:E1}), gives the energy of the ground
state as well as the excited state spectrum by using the Pauli principle
to assign
normal mode quantum numbers.
The Pauli allowed states are found by setting up a correspondence between
the states identified by normal mode quantum numbers
$|n_{{\bf 0}^+},n_{{\bf 0}^-},n_{{\bf 1}^+},n_{{\bf 1}^-},n_{\bf 2}>$ and the
non-interacting states of the trap with quantum numbers
$\nu_i$, the radial quantum number and $l_i$, the
 orbital angular momentum quantum number of the three dimensional harmonic
oscillator, $(V_{\mathtt{conf}}(r_i)=\frac{1}{2}m\omega_{ho}^2{r_i}^2)$. These
single-particle quantum numbers satisfy  $n_i = 2\nu_i + l_i$, where $n_i$ is the ith particle energy level quanta defined by: 
$E=\sum_{i=1}^N\left[n_i  +\frac{3}{2}\right] \hbar\omega_{ho} =
\sum_{i=1}^N \left[(2\nu_i + l_i) +\frac{3}{2}\right] \hbar\omega_{ho}$. 
The states of the harmonic oscillator have known constraints 
due to antisymmetry that can be transferred to the normal mode
representation in the double limit
  $D\to\infty$, $\omega_{ho}\to\infty$ where both
representations are valid.  The radial and angular
quantum numbers separate at this double limit giving
two conditions\cite{prl,harmoniumpra}:
\begin{equation} \renewcommand{\arraystretch}{1} 
\label{eq:quanta}
2 n_{{\bf 0}^-} + 2 n_{{\bf 1}^-} =   \sum_{i=1}^N 2 \nu_i \, ,\,\,\,
2 n_{{\bf 0}^+} + 2 n_{{\bf 1}^+} + 2 n_{\bf 2} = \sum_{i=1}^N  l_i  \,
\renewcommand{\arraystretch}{1}
\end{equation}
\noindent 

Eqs.~(\ref{eq:quanta}) define a possible set of normal mode states
$|n_{{\bf 0}^+},n_{{\bf 0}^-},n_{{\bf 1}^+},n_{{\bf 1}^-},n_{\bf 2}>$
 consistent with an antisymmmetric wave function
from the set of 
 harmonic oscillator configurations that are known to obey the Pauli principle.
As particles are added at the non-interacting
 $\omega_{ho}
\rightarrow \infty$ limit, additional harmonic oscillator quanta, $\nu_i$ 
and $l_i$, 
are, of course, required by the Pauli principle as fermions fill the
harmonic oscillator levels. Equivalently,
this corresponds to additional
 normal mode quanta required to ensure antisymmetry as 
the normal modes begin to
reflect the emerging interactions.
This strategy is analogous to  Landau's use of the non-interacting system 
in Fermi liquid theory to 
set up the correct Fermi statistics as interactions adiabatically evolve\cite{landau}.

\section{Application: Ultracold Fermi Gases from BCS to Unitarity}\label{sec:unitarygas}

I assume an $N$-body system of fermions with equal numbers of 
``spin up'' and ``spin down''  fermions
and $L=0$
symmetry. The particles are confined by a spherically-symmetric
harmonic potential with frequency $\omega_{ho}$ so 
$a_{ho}(=\sqrt{\hbar/(m\omega_{ho})})$ and $\omega_{ho}$  
 are the characteristic length and energy  
scales of the trap, representing the largest length scale  
and the smallest energy scale of the problem. An attractive square-well 
potential of radius $R$ is set up with a potential depth parameter 
$\bar{V_0}$ in scaled units which is varied from a value of 1.0 where the 
magnitude of the
s-wave scattering length, $a_s$, is infinite to a value of zero 
 as the gas becomes weakly interacting in the BCS regime. 
The range is chosen so $R \ll a_{ho}$.
(See also Ref.~\cite{prl} for a more detailed description of the potential.)


When the scattering length $a_s$ is
much smaller than the interparticle spacing the system is considered
weakly interacting.
To reach the strongly interacting unitary regime, a Feshbach resonance can 
be tuned using an external magnetic field so the scattering length
becomes much larger than the other length scales of the problem.
The system is strongly interacting in this regime and is independendent of
the microscopic details acquiring universal behavior.

I apply the full SPT many-body formalism defining the internal displacement 
coordinates
and determining the symmetry coordinates, the normal modes
and their frequencies as a function of $N$\cite{FGpaper,energy}.
 The energy expression of Eq.~(\ref{eq:E1}) gives the ground state energy
as well as the  excited state spectrum used in constructing the partition
function. I chose values of $N$ in the range $10 \le N \le 30$ which had
produced excellent results in the unitary regime. 
 For the thermodynamic quantities, converging
the partition function for higher values of $N$ becomes extremely difficult.

The canonical partition function is defined as: 
$Z = \sum_{j=0}^\infty g_j \exp(-E_j/T)$, where $E_j$ is a many-body energy,
$T$ is the temperature ($k_B=1$), and  $g_j$ 
is the degeneracy of $E_j$. To determine a particular degeneracy,
I search for all the partitions of the $N$ 
particles into different levels,  $n_i, i=1,,,N$ that yield the correct
$E_j$. For each partition, I find the possible 
quantum numbers
$l_i$ and $\nu_i$ of the occupied sublevels for all possible arrangements
of the particles. Gathering these statistics yields the degeneracy as well as the 
sums over $l_i$ and $\nu_i$ for this partition.
I then use Eq.~(\ref{eq:quanta}) to assign the normal mode quantum numbers
to ensure antisymmetry.
The quanta corresponding to
 the lowest  normal mode
frequencies are chosen 
yielding the lowest energy for each excited energy level. This gives 
occupation in $n_{\bf 2}$, the phonon modes, and in $n_{{\bf 1}^-}$, 
the particle-hole radial excitation modes, 
which have the lowest angular and radial  
frequencies 
respectively. The conditions are:
\begin{equation} \renewcommand{\arraystretch}{1} 
\label{eq:quanta3}
2 n_{{\bf 1}^-}  =  \sum_{i=1}^N 2 \nu_i ,\,\,\,\,\,\,\,
2 n_{\bf 2} = \sum_{i=1}^N  l_i \,. 
\renewcommand{\arraystretch}{1}
\end{equation}

Thus, the enforcement of the Pauli principle yields  
occupation in different normal
modes for each state determining the energy 
as well as character of the state since the normal modes have clear dynamical motions\cite{annphys}.

\subsection{Ground state energies from BCS to Unitarity}

\begin{figure}
\includegraphics[scale=0.5]{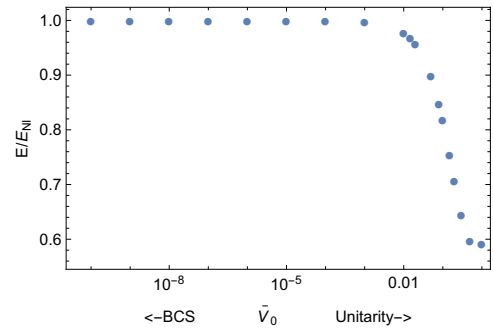}
\renewcommand{\baselinestretch}{0.8}
\caption{The SPT ground state energies from BCS to unitarity as a function 
of $\bar{V}_0$ for $N = 12$.}
\label{fig:one}
\end{figure}

Ground states energies  have been determined for trapped Fermi gases 
across the transition from BCS to unitarity using the SPT formalism.  
The SPT energies as a function of $\bar{V}_0$
are shown in Fig.~\ref{fig:one} from a value of $\bar{V}_0=10^{-8}$ deep in the
BCS regime to a value of $\bar{V}_0=1.0$ at unitarity. 
The energies are normalized by the 
noninteracting energies, $E_{NI}$, and increase rather rapidly from the values
at unitarity converging to the expected noninteracting energies, $E_{NI}$
as $\bar{V}_0 \rightarrow 0$. 
The energies at unitarity were determined in a previous 
study\cite{prl}
and compared to other theoretical values, agreeing closely with benchmark
auxilliary Monte Carlo results\cite{carlson} 
for $N \le 30$. (See Fig. 1 in Ref.~\cite{prl}.)

Unlike many approaches in the literature that use the s-wave scattering 
length, $a_s$,
 to set up  a contact interaction for the interparticle interaction, 
the SPT method does not explicitly use the scattering length 
to define the interaction term. (The square-well potential has a
scattering length associated with it, however, the solution of the perturbation
equations is only through first order, so the results reflect only the
first order terms from this potential, not the full scattering length.)
In order to compare to 
both experimental and theoretical results in the literature, I have 
used simple interpolation between my ground state energies across the
transition with ground state energies in the literature
that have been obtained using
an explicit scattering length in the interaction term. This connects the
interaction parameter $\bar{V}_0$ used in my SPT calculation 
to a value of the scattering length in a study using an explicit 
scattering length in the interaction
term.   (Since these two parameters have very different ranges
($ 0 \leq \bar{V}_0 \leq 1.0$; \,\, $-\infty \leq a_s \leq 0$; ) determining  
a scale factor between the parameters is probably not as accurate as
interpolation.)

\begin{figure}
\includegraphics[scale=0.6]{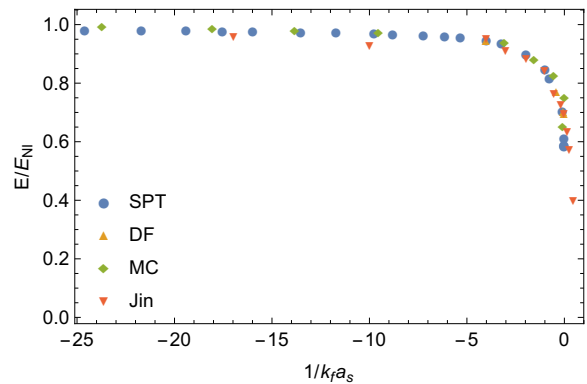}
\renewcommand{\baselinestretch}{0.8}
\caption{Ground state energies from BCS to unitarity as a function of 
$1/k_fa_s$. My SPT results are for $N = 12$ and are compared to 
experimental \cite{jin3}, density functional (DF)\cite{adhikari1}
and variational Monte Carlo results (MC)\cite{sanchez}}.
\label{fig:two}
\end{figure}

I chose to use the ground state energies from  a density functional
calculation\cite{adhikari1} which were obtained by fitting their 
interaction parameters to very accurate energies for the trapped superfluid 
both at unitarity\cite{blume3,chang} and in the BCS regime\cite{lee}. 
In Fig.~\ref{fig:two}, I have regraphed the SPT energies as a function of
these interpolated scattering lengths, specifically as a function
of $1/k_fa_s$ where $k_f$  is the Fermi momentum, and compared to 
available theoretical results\cite{sanchez} 
(including the density functional results
used for the interpolation\cite{adhikari1}) and
experimental results\cite{jin3}. For the experimental results which are for
potential energies across the transition, I have assumed that the virial
theorem which is valid at unitarity and at the independent particle limit
holds across the transition\cite{adhikari1,son}.
Using the results of other energy studies
across this transition for the interpolation 
yields comparable results as the close agreement
in Fig.~\ref{fig:two} would suggest.

\subsection{Entropies from BCS to Unitarity}

\begin{figure}
\includegraphics[scale=0.6]{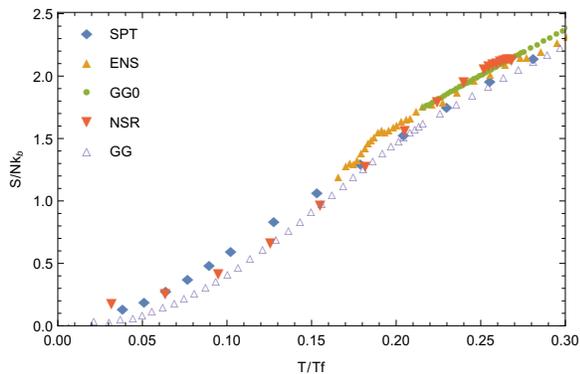}
\renewcommand{\baselinestretch}{0.8}
\caption{The SPT entropy for $N = 20$ as a function of $T/T_F$ at unitarity
is compared to experimental data: ENS\cite{nascimbene}
 and theoretical results: NSR, GG0, and GG\cite{hu1,hu2}.} 
\label{fig:three}
\end{figure}

\begin{figure}
\includegraphics[scale=0.6]{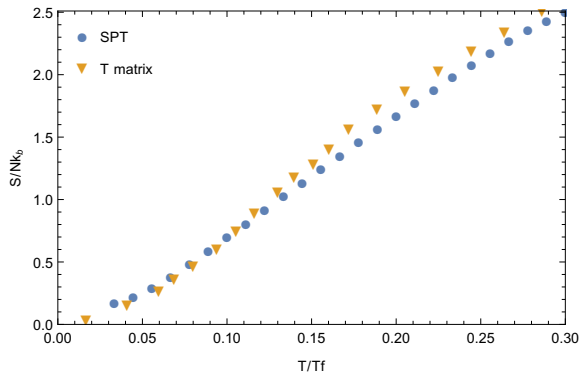}
\renewcommand{\baselinestretch}{0.8}
\caption{The entropy for $N = 20$ as a function of $T/T_F$  for 
$1/k_{\mathtt{f}a} =0.5$. SPT results - blue dots, T matrix results - orange
triangles from Ref.~\cite{levin2}}
\label{fig:four}
\end{figure}

Although thermodynamic quantities have been well studied in the unitary
regime, there are very few determinations of thermodynamic quantities
across the BCS to unitarity transition. I have chosen to look
at entropies across this transition since values for the entropy as a
function of temperature have been calculated at several values of  
$1/k_fa_s$ 
using a T-matrix approach\cite{levin2}.
My approach  uses a straightforward
calculation of the partition function, summing over the spectrum of
 equally-spaced normal mode states that are chosen by the
 Pauli principle. 
Using the interpolated values of the 
scattering length obtained above, I have plotted values for the entropy
at unitarity, $1/k_fa_s = 0$ ($a_s = -\infty$), in Fig.~\ref{fig:three} 
which agree well with theoretical\cite{hu1,hu2}
 and experimental results\cite{nascimbene} 
and at $1/k_fa_s = -0.5$ in Fig.~\ref{fig:four} 
comparing to the theoretical results of Ref.~\cite{levin2}.

The partition function becomes difficult to converge 
as the interparticle interaction decreases away from unitarity due
 to two effects: the narrowing of the frequencies and the
increase in the value of the frequencies as they approach $2\omega_{ho}$ 
deep in the BCS regime. The larger frequency
values mean that the individual terms of the partition function decrease their
contribution to the total (a larger negative number in the numerator of each 
exponential) so more states are needed for convergence. 
The narrowing of the frequencies as the gaps shrink toward the BCS regime
 means that more states are becoming accessible at a given temperature
which again increases the number of terms needed for convergence.
This increase in the number of states as interactions weaken
results in higher entropy values as can be seen in
Fig.~\ref{fig:four} for the weaker interactions at 
$1/k_fa_s = -0.5$ compared
to unitarity results in Fig.~\ref{fig:three}. The number of states needed also increases as the temperature 
increases. These three effects combine to make it very challenging to calculate
thermodynamic quantities accurately across the BCS to unitarity transition
using straightforward summing over the available states. 
Alternative approaches to obtaining a converged partition function
are complicated by the
need to enforce the Pauli principle at each step.

\subsection{Estimate of Critical Temperatures from BCS to Unitarity}

The critical temperature, $T_C$, is defined as the transition temperature 
from a normal fluid 
to a superfluid which exhibits long-range order due to a macroscopic occupation
of the phonon ground state.
This transition has been observed in the heat capacity whose
thermodynamic expression involves a derivative with respect to the 
temperature. The heat capacity
has a well-known, strong experimental signature in the unitary regime,
the lambda transition, 
that has been studied extensively both
experimentally\cite{thomas1,kinast1,thomas,thomas2,zwierlein3} and 
theoretically\cite{bulgac,burovski2,hu2,hu5}.
An estimate of the critical temperature in the unitary regime has also
been extracted from measurements of the
entropy as a function of temperature using the thermodynamic relation:
$1/T = \partial{S}/\partial{E}$\cite{thomas}. 

Theoretically, the sudden change in thermodynamic properties
 as the ensemble becomes a superfluid is governed by the partition function
and originates
in the details of lowest terms including the size of the gap and the
degeneracies of the lowest states.  For a given spectrum, the partitioning of 
particles among the
available energy levels depends on a single parameter, the temperature.
As the temperature drops below the critical temperature, one expects to see
 the occupation in the phonon ground state increase rapidly
due to the gap in the spectrum. 
This phenomenon is manifested by 
a sudden change in the value of certain observables such as the
specific heat.

\begin{figure}
\includegraphics[scale=0.6]{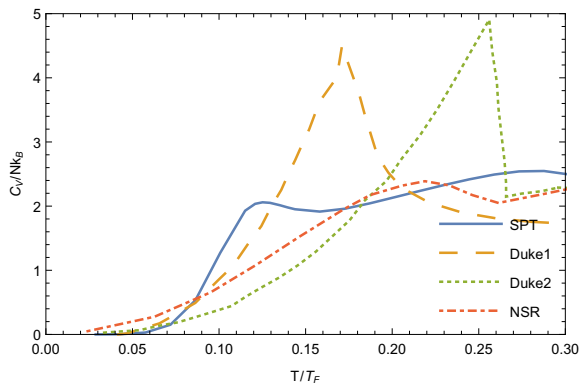}
\renewcommand{\baselinestretch}{0.8}
\caption{The heat capacity showing with cusp at the critical temperature, 
$T_C/T_F$ at unitarity comparing to experimental: Duke1\cite{thomas1},
Duke2\cite{kinast1}
and theoretical results: NSR\cite{hu1} as a function of temperature.}
\label{fig:five}
\end{figure}

In an earlier SPT study in the unitary regime\cite{emergence},  
a calculation of the specific
heat clearly showed a cusp at the lambda transition, 
yielding a critical temperature of  $(T/T_F)_C=0.16$ 
which was significantly
 lower than previous results in 
the literature for trapped Fermi gases:
$(T/T_F)_C=0.19$\cite{nascimbene}, $0.20$\cite{burovski2},
 $0.21$\cite{hu2,hu5,haussmann3}, 
$(T/T_F)_C=0.27$\cite{kinast1,hu5,bulgac}, 
0.29\cite{thomas,hu5}. (See Fig.~\ref{fig:five} reproduced from Fig. 12
in Ref.~\cite{emergence}.) 

\begin{figure}
\includegraphics[scale=0.6]{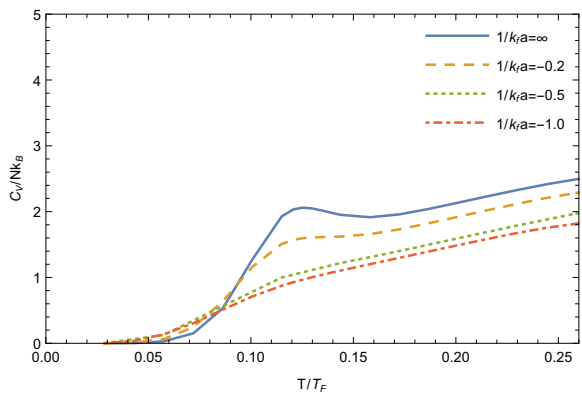}
\renewcommand{\baselinestretch}{0.8}
\caption{The heat capacity showing the softening of the cusp as the 
interparticle interaction decreases from a maximum at unitarity:
blue line $1/k_fa = \infty$; orange dashed line $1/k_fa = -0.02$; 
green dotted line $1/k_fa = -0.5$; red dot dashed line $1/k_fa = -1.0$.}
\label{fig:six}
\end{figure}

I have repeated this calculation for weaker interactions, $1/k_fa = -0.02$,
$1/k_fa = -0.5$, and $1/k_fa = -1.0$,
 graphing the results in Fig.~\ref{fig:six}.
As the interactions become weaker, the excitation gap decreases and the
cusp signifying a transition to a superfluid quickly softens. While still
visible at  $1/k_fa = -0.02$ close to the unitary limit, the cusp
is undetectable at a value of
$1/k_fa \leq -0.5$ in the crossover region with just a slight
inflection visible, and by $1/k_fa = -1.0$ no sign is detected. 
Thus, observing an experimental signature of this transition, 
certainly a definitive 
way to define the critical temperature, is not always possible in all
regimes.

Theoretically, several approaches have been used to estimate the critical
temperature at unitarity including a Monte Carlo 
study\cite{burovski,burovski2} that uses the behavior of 
 a correlation function to determine an estimate of the critical 
temperature, and an auxiliary field quantum Monte Carlo approach that
determines the critical temperature from a change in the behavior of
the thermodynamic energy as a function of temperature\cite{bulgac}.
Along the entire transition from BCS to unitarity, the critical temperature
has been calculated by solving the gap equation self-consistently with the 
number equation under the condition that the order parameter
goes to zero as the temperature approaches the critical temperature, $T_C$, 
from below, i.e. long-range order
is lost.  These equations have
been solved at different
levels of approximation from mean field which yields the well known BCS results
to solutions in strongly interacting regimes near unitarity that include full 
fluctuations\cite{grava,strinati1,melo}. 

\begin{figure}
\includegraphics[scale=0.6]{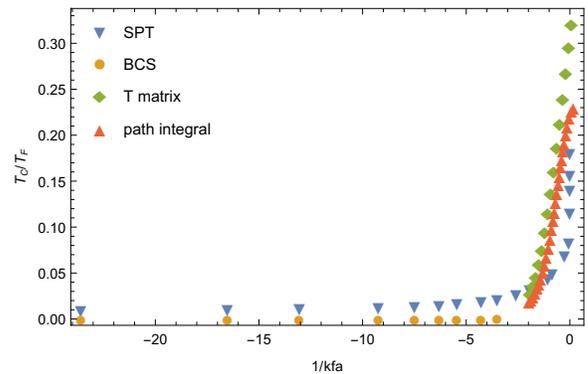}
\renewcommand{\baselinestretch}{0.8}
\caption{The critical temperature, $T_C/T_F$ as a function of 
$1/k_fa$ for 30 fermions compared to the BCS prediction valid 
for $1/k_fa \ll -1.0$
and theoretical results\cite{strinati1,grava} for $-1.0 \leq 1/k_fa \ll 0$.}
\label{fig:seven}
\end{figure}

The SPT approach offers an alternative, straightforward  way to estimate
 the critical temperature across the entire transition.
Using the Pauli principle, the first excited state 
above the ground
state can be determined along the transition. This excited
state involves single particle excitations while the ground state is
 composed only
of phonon normal modes. 
The difference between these two states gives an estimate of the critical 
temperature as
simply the temperature equivalent: $E_{ex}-E_{gs} = k_BT_C$.  This estimate
is graphed in Fig.~\ref{fig:seven} normalized by the Fermi temperature $T_F$,
$E_F=(3N)^{1/3}\hbar\omega_{ho}=k_BT_F$, and compared to other 
theoretical results in the region near unitarity and 
to the BCS expression, $T_C/T_F = 0.277 \exp(\pi/(2k_fa)$ valid
for $1/k_fa \ll -1.0$. The SPT
results are slightly higher than the BCS results, showing
 a gradual increase from the deep BCS regime toward unitarity and 
then a rapid increase
for $1/k_fa \ge -1.0$ as the interactions approach unitarity.
The curve converges at unitarity at  $(T/T_F)_C=0.18$ in reasonable
 agreement with
several other theoretical 
approaches\cite{nascimbene,burovski2,hu2,hu5,haussmann3}

\subsection{The breathing mode frequency from BCS to 
Unitarity.}

Investigating collective excitation
modes has long been used to gain insight into the
behavior of many-body systems.  The excitation frequencies of ultracold
Fermi gases have been studied intensely across the BEC-BCS transition.
The radial compression or ``breathing'' mode in
a cylindrical potential has been of particular interest due to a 
surprising feature observed
in the regime of strong interactions, specifically an abrupt
decrease in the frequency near unitarity\cite{thomas1,bartenstein,grimm1,grimm2}.  
This minimum
has been confirmed theoretically\cite{tosi,zubarev,salasnich}.

The microscopic basis for this minimum in the breathing mode has been
attributed to the formation of Cooper pairs as unitarity is approached which
decreases the frequency as the gas becomes more compressible\cite{tosi}.
It has also been suggested from the observation of this minimum
 coupled with an analysis of the corresponding damping time, that this
feature could be a signature of a transition from a superfluid to
a collisionless phase\cite{bartenstein,thomas1} as interactions weaken
toward the independent particle regime.

In a recent paper, SPT was used to study 
the five analytic normal mode excitation frequencies  in a symmetric trap
across the transition
from BCS to unitarity\cite{prafreq}.  This study
revealed that unlike the angular frequencies which converged smoothly
to integer multiples of the trap frequency, 
the two radial frequencies went through a minimum as
unitarity is approached and then continued to increase suggesting that
these first order frequencies might not be
fully converged at unitarity. (See Fig. 2 in Ref.~\cite{prafreq}.)

\begin{figure}
\includegraphics[scale=0.6]{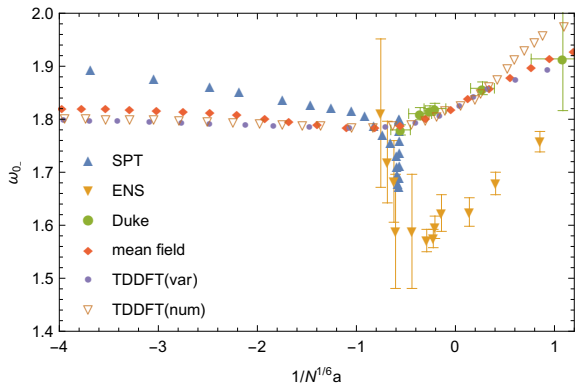}
\renewcommand{\baselinestretch}{0.8}
\caption{Excitation frequency for the radial breathing mode in a symmetric
trap as a function 
of $1/N^{1/6}a$ for 30 fermions showing the minimum as unitarity is 
approached
that is seen theoretically\cite{zubarev,salasnich,tosi} and 
experimentally in cylindrical traps\cite{grimm2,thomas1}}
\label{fig:eight}
\end{figure}

In  Fig.~\ref{fig:eight}, I have regraphed in the SPT radial 
breathing normal mode frequency 
$\omega_{{\bf 0}^-}$ as a
function of the parameter $1/N^{1/6}a$ through the region of the minimum 
and compared to these earlier results.
Fig.~\ref{fig:eight} clearly shows a minimum  for the SPT frequency between
$1/N^{1/6}a = -1.0$ and unitarity, $1/N^{1/6}a = 0$, 
in close 
agreement with these previous
experimental and theoretical 
results\cite{grimm2,thomas1,tosi,zubarev,salasnich}.
The SPT minimum is broad in Fig. 2 in Ref.~\cite{prafreq} graphed
as a function of $\bar{V}_0$ on a log scale spanning several orders of 
magnitude from
$\bar{V}_0 = 10^{-3}$ to $\bar{V}_0 =1.0$, but is quite sharp when graphed as 
a linear function of  $-1/N^{1/6}a$ in Fig.~\ref{fig:eight} where it maps into 
a small region 
between $1/N^{1/6}a = -1.0$ and $1/N^{1/6}a = 0$.

The other SPT radial excitation, $\omega_{{\bf 1}^-} $, which is a 
single-particle excitation also clearly
shows a minimum when graphed as a function of $\bar{V}_0$ in Fig. 2 of 
Ref.~\cite{prafreq}.
When regraphed as a function of $1/N^{1/6}a$, this minimum is visible, but
quite small.

The analytic form of the SPT normal modes offers an opportunity to
analyze the microscopic dynamics responsible for this minimum.
 By tracking the contribution of different
terms in the Hamiltonian to the analytic expression for the frequency
across the transition, one can understand what is happening microscopically
to produce this minimum.
The following analysis is based on previous work in
Ref~\cite{prafreq}.

\subparagraph{Understanding the microscopic dynamics of the minimum in
the radial breathing frequency.}

The analytic expressions for the frequencies have been 
studied across the transition from BCS to unitarity\cite{prafreq}.
Appendix C in Ref.~\cite{prafreq} has an analysis of the radial breathing
frequency
$\omega_{{\bf 0}^-}$ in terms of the $FG$ matrix elements from the terms
in the first-order Hamiltonian (Eq.~(\ref{eq:H0y})). 
The formula derived in this Appendix for
$\omega_{{\bf 0}^-}$ in terms of the $FG$ elements is:

\begin{equation}
\omega_{{0}^{-}}  \approx  \sqrt{G_aF_a+(N-1)G_aF_b} \label{eq:omega0mFG} \\
\end{equation}

\noindent where $G_a =1$, $F_a$ and $F_b$ involve derivatives of 
 $\bar{V}_{\mathtt{eff}}$ (See Eq.~\ref{eq:veff_BEC}.) which is a sum of
the confining potential $\bar{V}_{\mathtt{conf}}$, 
the centrifugal potential $\bar{V}_{\mathtt{cent}} = \bar{U}$, 
and the interparticle interaction
potential $\bar{V}_{\mathtt{int}}$:

\begin{equation} \label{eq:Veff}
\bar{V}_{\mathtt{eff}} =
\bar{U}+\bar{V}_{\mathtt{conf}}+\bar{V}_{\mathtt{int}}.
\end{equation}

\noindent yielding three terms for $F_a$:
$F_a=  F_{a}^{\mathtt{cent}} + F_{a}^{\mathtt{conf}} +F_{a}^{\mathtt{int}}$ 
and one nonzero term
for $F_b$ involving the interaction potential: 
$F_b= F_{b}^{\mathtt{int}}$.
The term $F_{a}^{\mathtt{conf}}$ is a constant equal to 1. 
All the  terms are explicitly
defined in Appendix B in Ref.~\cite{prafreq}.

As in the analysis of the angular frequencies in Section VI of 
Ref.~\cite{prafreq}, it
is useful to track the magnitude of $\gamma_{\infty}$, the angle cosine
of each pair of particles at the minimum of the maximally-symmetric
structure at large dimension. Early dimensional
scaling work identified a nonzero value of this parameter as a signature of 
the existence of correlation between the particles.
Mean field results have  $\gamma_{\infty} =0$ corresponding to 
no correlation between the particles, while increasing values of 
 $\gamma_{\infty}$ indicated stronger and longer-range correlation effects 
existed.

Consider the independent particle limit, i.e. collisionless regime, 
with no interparticle interactions so
$\bar{V}_0 =0$ and thus no correlations between the particles i.e.
$\gamma_{\infty} = 0$ so only the harmonic trap is affecting the
particles which of course are also obeying the Pauli principle.
Most of the terms in the
expression for  $\bar{\omega}_{{0}^{-}}$ in Eq.~(\ref{eq:omega0mFG}) are 
zero.  The only non-zero terms
are $F_{a}^{\mathtt{conf}} =1$ from the trap potential and $F_{a}^{\mathtt{cent}} =3$ 
which originates in the kinetic energy, giving
$F_{a} =4$, $F_{b}=0$ so  $\omega_{{\bf 0}^-} = 2\omega_{ho}$ 
as expected and confirmed in the laboratory. 
(See Appendix F in Ref.~\cite{prafreq}.) 
 As interactions are introduced, 
$\gamma_{\infty}$ assumes a small
nonzero value, signaling the existence of weak correlations. This nonzero
value means that all of the terms in the expression for $\bar{\omega}_{{0}^{-}}$
 are nonzero.
$F_{a}^{\mathtt{cent}}$ begins to decrease, while $F_{a}^{\mathtt{int}}$ and 
$F_{b}^{\mathtt{int}}$ increase. 
Along the BCS to unitarity transition, the value of
 $\bar{\omega}_{{0}^{-}}$ is a balance between the centrifugal
term which is decreasing and the interaction terms which are increasing 
as interactions ($\bar{V}_0$) and correlations ($\gamma_{\infty}$) 
both increase from BCS toward unitarity. The minimum in
the frequency occurs from the continued decrease in the centrifugal terms just
before the increase in the interaction terms dominates.

Microscopically, one can understand what is happening from this analysis
of the Hamiltonian terms. The increase in the correlated motion of the 
particles as tracked by the 
increase in $\gamma_{\infty}$ minimizes the interparticle interactions
resulting in slower oscillations of the breathing mode. Eventually the
increase in $\bar{V}_0$, i.e. the increased strength of the
 interparticle interactions
will, of course, lead to more rapid oscillations i.e. 
an increase in the frequency as unitarity is 
approached.  The gradual decrease seen when  $\bar{\omega}_{{0}^{-}}$ is 
graphed as a function of $\bar{V}_0$ in Ref.~\cite{prafreq}
appears as a sudden, quite narrow dip in the frequency when graphed as a 
function of $1/(N^{1/6}a)$. This is due to the rapidly changing
 scattering length in this region
as unitarity is approached. In summary, the minimum is the result of two
competing factors which affect the microsopic behavior: 
the increase in correlation which minimizes the interparticle interactions
thus slowing down the
frequency of the oscillations and second, 
the increasing strength of the interparticle interactions 
which eventually dominates and speeds up the frequency of the oscillations.

\section{Summary and Conclusions} \label{sec:SumConc}

In this study, I have explored the ability of normal modes
to describe the behavior of ultracold Fermi gases including superfluidity 
across the BCS to unitarity transition
without assuming Cooper pairing. In particular, I have calculated the
following properties: ground state energies, thermodynamic entropies, 
critical temperatures and the radial
breathing frequency across this transition using normal modes
and compared to available  
experimental and theoretical results. 

An earlier study that looked at the behavior of the SPT normal mode 
frequencies across this transition found that the 
 frequencies were capable of describing the emergence of
excitation gaps from very small gaps deep in the BCS regime
to a maximum at unitarity as observed experimentally. In addition,
they provided insight
into the microscopic dynamics responsible for the universal behavior
at unitarity.
This was possible due to the analytic forms of both the $N$-body 
normal mode 
frequencies and their coordinates.

The success of this earlier study motivated the current exploration of 
additional
properties across the transition. This study 
has yielded close agreement with both experimental and theoretical results
for the ground state energies, thermodynamic entropies and critical 
temperatures
at weaker interactions away from unitarity. These calculations tested
 the lowest frequencies
relevant to ultracold systems as well as the spectrum of frequencies needed
for the partition function. In all
these calculations, the Pauli principle has played a central role
in choosing the
states that contribute to these properties. The final
study involved a single frequency, the  breathing frequency, which
did not contribute to the earlier studies due to its larger value.
The observed dip in this SPT frequency near unitarity was in close
agreement with results first observed
in the laboratory and later confirmed theoretically, suggesting that 
the first-order SPT Hamiltonian that produces this frequency contains 
sufficient physics to describe this transition.

The normal coordinates constitute 
beyond-mean-field, analytic solutions to a many-body Hamiltonian
and offer insight microscopically into the evolution of 
properties
across the BCS to unitarity transition.
The analytic forms for the frequencies and coordinates allow a detailed
look at the dynamics by tracking the  effect of Hamiltonian terms 
across the transition.
As correlations increase toward unitarity as 
tracked by the parameter $\gamma_\infty$, the
dependence of properties on the details of the
interparticle interactions is minimized consistent with the universal behavior
which is also seen at the independent
particle limit. The Pauli principle, of course, is dominating the dynamics
at both limits underpinning the universal behavior in these regimes.

The results of this study are based on an  exact solution 
of the first-order equation of SPT perturbation theory which contains
beyond-mean-field effects. Higher order terms which have
been formulated, but not implemented, are not included. 
These terms could become significant
in some regimes changing the dynamics.
The SPT formalism also does not offer a mechanism for the 
two-body pairing 
that occurs as the ensemble transitions to the BEC regime.

The successful description of these properties across the BCS to unitarity
transition using these first-order normal
mode solutions supports my earlier conclusions that normal
modes are a viable model of superfluidity across a broad range of interparticle
interaction strengths and represent an interesting alternative to 
Cooper pairing models.

\section{Acknowledgments}
I am grateful to the National Science Foundation for financial support
under Grant No. PHY-2011384.

\bigskip

\appendix
\renewcommand{\theequation}{A\arabic{equation}}
\setcounter{equation}{0}

\end{document}